\documentclass[prl,12pt,showpacs]{revtex4}

\usepackage{dcolumn}
\usepackage{amssymb}
\usepackage{amsmath}
\usepackage{graphicx}

\begin{document}

\preprint{APR 2004-XX}

\title{Spin-Dependent Transport of Electrons in a Shuttle Structure}

\author{L. Y. Gorelik}
\email{gorelik@fy.chalmers.se}
\affiliation{Department of Applied
Physics, Chalmers University of Technology and G\"{o}teborg
University, SE-412 96 G\"{o}teborg, Sweden}
\author{S. I. Kulinich}
\affiliation{Department of Applied Physics, Chalmers University of
Technology and G\"{o}teborg University, SE-412 96 G\"{o}teborg,
Sweden}
\affiliation{B.~I.~Verkin Institute for Low Temperature
Physics and Engineering, 47 Lenin Avenue, 61103 Kharkov, Ukraine}
\author{R. I. Shekhter}
\affiliation{Department of Applied Physics, Chalmers University of
Technology and G\"{o}teborg University, SE-412 96 G\"{o}teborg,
Sweden}
\author{M. Jonson}
\affiliation{Department of Applied Physics, Chalmers University of
Technology and G\"{o}teborg University, SE-412 96 G\"{o}teborg,
Sweden}
\author{V.~M.~Vinokur}
\affiliation{Materials Science Division, Argonne National
Laboratory, 9700 South Cass, Argonne, Illinois 6043}

\date{\today}

\begin{abstract}
We consider "shuttling" of spin-polarized electrons between two
magnetic electrodes (half-metals) by a movable dot with a single
electronic level. If the magnetization of the electrodes is
antiparallel we show that the transmittance of the system can be
changed by orders of magnitude if an external magnetic field,
perpendicular to the polarization of the electronic spins, is
applied. A giant magnetotransmittance effect can be achieved for
weak external fields of order $1\div 10$ Oe.
\end{abstract}


\maketitle

\section{Introduction}
Metal-organic nanocomposite materials are interesting from the
point of view of the "bottom-up" approach to building future
electronic devices. The ability of the organic parts of the
composite materials to identify and latch on to other organic
molecules is the basis for the possible self assembly of nanoscale
devices, while the metallic components provide mechanical
robustness and improve the electrical conductance.

Such composite materials are heteroelastic in the sense that the
mechanical rigidity of the organic and metallic components are
very different. This allows for a special type of deformation,
where hard metallic components embedded in a soft organic matrix
can be rearranged in space at a low deformation energy cost
associated with stretching and compressing the soft matrix. Strong
Coulomb forces, due to accumulation of electronic charge in
embedded nanoscale metallic particles, can be a source of such
mechanical deformations. This leads to a scenario where the
transport of electric charge, possibly due to tunnelling of
electrons between metal particles, becomes a complex
nano-electromechanical phenomenon, involving an interplay of
electronic and mechanical degrees of freedom \cite{1}. Such an
interplay can lead to new physics, as was recently demonstrated
theoretically for the simplest possible structure --- a
Nanoelectromechanical single-electron transistor. The
electromechanical instability predicted to occur in this device at
large enough bias voltage was shown to provide a new mechanism of
charge transport \cite{2}. This mechanism can be viewed as a
''shuttling" of single electrons by a metallic island
--- a Coulomb dot --- suspended between two metal electrodes. The
predicted instabilily leads to a periodic motion of the island
between the electrodes shuttling charge from one to the other.

The shuttle instability appears to be a rather general phenomenon.
It has, {\em e.g.}, been shown to occur even for extremely small
suspended metallic particles (or molecules) for which the coherent
quantum dynamics of the tunnelling electrons \cite{3} or even the
quantum dynamics of the mechanical vibration \cite{4,5,6,7} become
essential. Nanomechanical transport of electronic charge can,
however, occur without any such instability, {\em e.g.}, in an
externally driven device containing a cantilever vibrating at
frequencies of order 100 MHz. A small metallic island attached to
the tip of the vibrating cantilever may shuttle electrons between
metallic leads as has recently been demonstrated \cite{8}. Further
experiments with magnetic and superconducting externally driven
shuttles as suggested in \cite{9}, seem to be a natural extension
of this work. Fullerene-based nanomechanical structures \cite{10}
are also of considerable interest.

The possibility to place transition-metal atoms or ions inside
organic molecules introduces a new "magnetic" degree of freedom
that allows the electronic spins to be coupled to mechanical and
charge degrees of freedom \cite{11}. By manipulating the
interaction between the spin and external magnetic fields and/or
the internal interaction in magnetic materials, spin-controlled
nanoelectromechanics may be achieved. An inverse phenomenon
--- nanomechanical manipulation of nanomagnets --- was suggested
earlier in \cite{011}. A magnetic field, by inducing the spin of
electrons to rotate (precess) at a certain frequency, provides a
clock for studying the shuttle dynamics and a basis for a dc
spectroscopy of the corresponding nanomechanical vibrations.

A particularly interesting situation arises when electrons are
shuttled between electrodes that are half-metals. A half-metal is
a material that not only has a net magnetization as do
ferromagnets, but all the electrons are in the same spin state
--- the material is fully spin-polarized. Examples of such
materials can be found among the perovskite maganese oxides, a
class of materials that show an intrinsic, so called "colossal
magnetoresistance" effect at high magnetic fields (of order 10-100
kOe) \cite{12}.

A large magnetoresistance effect at lower magnetic fields has been
observed in layered tunnel structures where two thin perovskite
manganese oxide films are separated by a tunnel barrier
\cite{12,Sun1,Sun2,Lu,13}. Here the spin polarization of
electronic states crucially affects the tunnelling between the
magnetic electrodes. This is because electrons that can be
extracted from the source electrode have there spins aligned in a
definite direction, while electrons that can be injected into the
drain electrode must also have there spins aligned --- possibly in
a different direction. Clearly the tunnelling probability and
hence the resistance must be strongly dependent on the relative
orientation of the magnetization of the two electrodes. An
external magnetic field aligns the magnetization direction of the
two films at different field strengths, so that the relative
magnetization can be changed between high- and low resistance
configurations. A change in the resistance of trilayer devices by
factors of order 2-5 have in this way been induced by magnetic
fields of order 200 Oe \cite{Sun1,Sun2,Lu}. The required field
strength is determined by the coercivities of the magnetic layers.
This makes it difficult to use a tunneling device of the described
type for sensing very low magnetic fields. In this paper we
propose a new functional principle --- spin-dependent shuttling of
electrons --- for low-magnetic field sensing purposes. We will
show that this principle can lead to a giant magnetoresistance
effect in external fields as low as 1-10 Oe.

The new idea which we propose to pursue is to use the external
magnetic field to manipulate the \textit{spin of shuttled
electrons} rather than the magnetization of the leads. The
possibility to "trap" electrons on a nanomechanical shuttle
(decoupled from the magnetic leads) during quite a long time on
the scale of the time it takes an electron to tunnel on/off the
shuttle makes it possible for even a weak external field to rotate
the electron's spin to a significant degree. Such a rotation
allows the spin of an electron, loaded onto the shuttle from the
spin-polarized source electrode, to be reoriented in order to
allow the electron finally to tunnel from the shuttle to the
spin-polarized drain lead. As we will show below, the magnetic
field induced spin-rotation of shuttled electrons is a very
sensitive nanomechanical mechanism for a giant magnetoresistance
(GMR) effect.

\begin{figure}
\vspace{-0.5cm} \centerline{
\includegraphics[width=20cm]{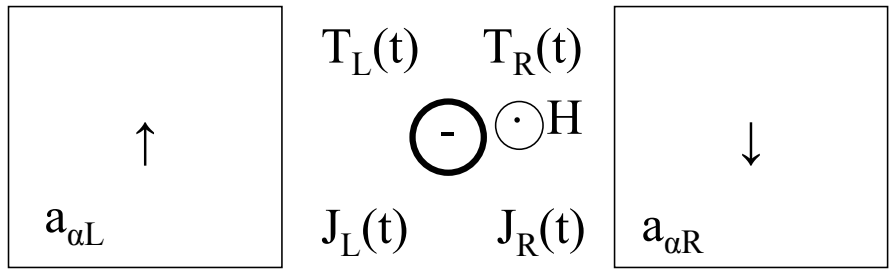} } \vspace*{-9cm}
\caption{Schematic view of the nanomechanical GMR device: a
movable dot with a single electron level couples to the leads due
to tunnelling of electrons, described by the tunnelling
probability amplitudes $T_{L,R}(t)$), and due to the exchange
interaction whose strength is denoted by $J_{L,R}(t)$. An external
magnetic field $H$ is oriented perpendicular to the direction of
the magnetization in the leads (arrows).}
\end{figure}

\section{ Formulation of the Problem. General Expression for
the Current} A schematic view of the nanomechanical GMR device to
be considered is presented in Fig.~1. Two fully spin-polarized
magnets with fully spin-polarized electrons serve as source and
drain electrodes in a tunnelling device. In this paper we will
consider the situation when the electrodes have exactly opposite
polarization. A mechanically movable quantum dot (described by a
time-dependent displacement $x(t)$), where a single energy level
is available for electrons, performs forced harmonic oscillations
with period $T = 2\pi/\omega$ between the leads. The external
magnetic field is perpendicular to the orientation of the
magnetization in both leads.

The Hamiltonian that governs the dynamical evolution of the system
is
\begin{eqnarray}\label{1}
 {\cal \hat H}(t)&=&\varepsilon_0(a^\dag_\uparrow a_\uparrow +
a^\dag_\downarrow a_\downarrow ) +\sum_\alpha( \varepsilon_\alpha
a^\dag_{\alpha,L} a_{\alpha,L}+\varepsilon_\alpha
a^\dag_{\alpha,R} a_{\alpha,R})\\&-& J_L(t)(a^\dag_\uparrow
a_\uparrow - a^\dag_\downarrow a_\downarrow)- J_R(t)
(a^\dag_\downarrow a_\downarrow - a^\dag_\uparrow a_\uparrow)-(g
\mu H/2)(a^\dag_\uparrow a_\downarrow +a^\dag_\downarrow
a_\uparrow) \nonumber \\ &+& T_L(t)\sum_\alpha (a^\dag_{\alpha,L}
a_\uparrow+a^\dag_\uparrow a_{\alpha,L})+ T_R(t)\sum_\alpha (
a^\dag_{\alpha,R} a_\downarrow+a^\dag_\downarrow
a_{\alpha,R})\,,\nonumber
\end{eqnarray}
where $a^\dag_{\alpha, L(R)}, (a_{\alpha, L(R)})$ are the creation
(annihilation) operators of electrons with the energy
$\varepsilon_\alpha$ on the left (right) lead (we have suppressed
the spin indices for the electronic states in the leads due to the
assumption of full spin polarization),
$a^\dag_{\uparrow(\downarrow)} (a_{\uparrow(\downarrow)})$ are the
creation (annihilation) operators on the dot, $\varepsilon_0$ is
the energy of the on-dot level, $J_{L(R)}(t)\equiv J_{L(R)}(x(t))$
are the exchange interactions between the on-grain electron and
the left (right) lead, $\Lambda_{L(R)}(t)\equiv
\Lambda_{L(R)}(x(t))$ are the tunnel coupling amplitudes, $g$ is
the gyromagnetic ratio and $\mu$ is the Bohr magneton.

The single-electron density matrix describing electronic transport
between the leads may be presented in the form:
\begin{equation}\label{5}
\hat\rho=\sum_\alpha
w_{\alpha,L}|\Psi^{\alpha,L}\rangle\langle\Psi^{\alpha,L}| +
\sum_\alpha
w_{\alpha,R}|\Psi^{\alpha,R}\rangle\langle\Psi^{\alpha,R}|\,.
\end{equation}
Here $|\Psi^{\alpha,L}\rangle$ are single-electron
states that obey the time-dependent Shr\"{o}dinger ($\hbar=1$)
equation with a Hamiltonian given by Eq.~(\ref{1}). The initial
condition has the form $$ |\Psi^{\alpha,L(R)}(t\rightarrow
-\infty)\rangle=|\alpha,L(R)\rangle \exp (-i\varepsilon_\alpha
t)\,,$$ where $|\alpha,L(R)\rangle$ is a single-electron state on
the left (right) lead with energy $\varepsilon_\alpha$.

We will suppose that the internal relaxation in the leads is fast
enough to lead to equilibrium distributions of the electrons. This
means that
$w_{\alpha,L(R)}=f(\varepsilon_\alpha \mp V/2)$ (where
$f(\varepsilon)$ is the Fermi distribution function) and $V$ is
applied voltage.

The problem at hand is greatly simplified if one considers the
large bias-voltage limit
\begin{equation}\label{01}
\mid V- \varepsilon_0\mid\gg \nu \Lambda_{\text{max}}^2\,,
\end{equation}
where $\nu$ is the density of states on the leads. The
restriction(\ref{01}) does not allow us to consider a narrow
transition region of voltages from the zero-current regime at $V <
\varepsilon_0 - \nu \Lambda_{\text{max}}^2$ to the fully
transmissive (in the absence of spin polarization effects) regime
at $V > \varepsilon_0 + \nu \Lambda_{\text{max}}^2$. However, it
covers the in practise most important case when the fully
transmissive junction is strongly affected by electronic
spin-polarization. Therefore, in our further considerations we
will take $w_{\alpha,L}=1$, $w_{\alpha,R}=0$ and $\varepsilon_{0}=
0$.

We will calculate the average current, $I$, through the system
from the relation
\begin{eqnarray}\label{50}
&&\hspace{.5in}I=\frac{1}{T}\int^T_0 dt \,\text{Tr}\,\{\hat \rho\hat\jmath\}\,,\\
&&\hat\jmath=e\frac{\partial\hat N_R}{\partial t}= i e [ \hat{\cal
H},\hat N_R]= i e T_R(t)\sum_\alpha(a^\dag_\downarrow a_{\alpha,R}
-a^\dag_{\alpha,R} a_\downarrow)\,,\nonumber
\end{eqnarray}
where $ \hat N$ is the electron number operator for the right
lead, $\hat N =\sum_\alpha a^\dag_{\alpha,R}a_{\alpha,R}$.

In general, the state $|\Psi^{\alpha,L}\rangle$ can be expressed
as
\begin{equation}\label{3}
|\Psi^{\alpha, L}(t)\rangle=c^\alpha_\uparrow (t)|\uparrow\rangle+
c^\alpha_\downarrow(t)|\downarrow\rangle
+\sum_\beta(c^{\alpha,\beta}_L(t)|\beta,L\rangle +
c^{\alpha,\beta}_R(t)|\beta,R\rangle)\,,
\end{equation}
Thus the problem is reduced to determining the coefficients
$c^{\alpha,\beta}_{R(L)}$ and $c^\alpha_{\downarrow(\uparrow)}$.

At this point it is convenient to introduce the bi-vectors
$$
\textbf{c}^{\alpha}=\left(\begin{array}{c} c^\alpha_\uparrow \\
c^\alpha_\downarrow \
\end{array}\right), \,\,\textbf{e}_1=\left(\begin{array}{c} 1 \\ 0 \
\end{array}\right), \,\,\mbox{and}\,\,\,\,\textbf{e}_2=\left(\begin{array}{c}
0 \\ 1 \
\end{array}\right)\,,$$
so that the coefficients $c^{\alpha,\beta}_{R(L)}$ can be
expressed as (see Appendix 1)
\begin{eqnarray}\label{10}
&c^{\alpha,\beta}_L&=e^{-i\varepsilon_\beta t}\delta_{\alpha\beta}
-i\int^t_{-\infty}dt' e^{i\varepsilon_\beta(t-t')}T_L(t')
(\textbf{e}_1,\textbf{c}^\alpha(t'))\,,\nonumber \\
&c^{\alpha,\beta}_R&=-i\int^t_{-\infty}dt'
e^{i\varepsilon_\beta(t-t')}T_R(t')
(\textbf{e}_2,\textbf{c}^\alpha(t'))\,.\nonumber
\end{eqnarray}
Here $(\textbf{a},\textbf{b})$ is the inner product of two
bi-vectors. As shown in Appendix 1, by using the wide band
approximation ({\em i.e.} by taking the electron density of states
in the leads $\nu$ to be constant) the equation for the bi-vectors
$ \textbf{c}^{\alpha}$ takes the form
\begin{equation}\label{9}
i\frac{\partial \textbf{c}^{\alpha}}{\partial t} = \hat R(t)
\textbf{c}^{\alpha} +\textbf{f}^\alpha(t) \,.
\end{equation}
Here $\textbf{f}^\alpha(t) = T_L(t)e^{-i\varepsilon_\alpha
t}\textbf{e}_1$ and the matrix $\hat{R}(t)$ is
\begin{equation}\label{8}
\hat R(t)=\left(\begin{array}{cc} - J(t)- i \Gamma_L(t)/2 &-g\mu
H/2
\\ -g\mu H/2 & J(t)-i \Gamma_R(t)/2 \
\end{array}\right)\,,
\end{equation}
where $J(t)=J_L(t)-J_R(t)$ and $\Gamma_{L(R)} (t)= 2\pi \nu
\Lambda_{L(R)} ^2(t)$ is the level width.

The formal solution of Eq.~(\ref{9}) can be written in the form
\begin{equation}\label{12}
\textbf{c}^{\alpha}(t)=-i\int^t_{-\infty}dt'\hat
L(t,t')\,\textbf{f}^{\alpha}(t')\,,
\end{equation}
where the "evolution" operator $\hat L(t,t'), (\hat L(t,t)=\hat
I)$, is defined as the solution of the equation
\begin{equation} \label{110}
i\frac{\partial\hat L(t,t')}{\partial t}= \hat R(t)\hat L(t,t')\,,
\end{equation}
and obeys the  multiplicative and periodicity properties,
\begin{equation}\label{120}
\hat L(t,t')=\hat L(t,t'')\hat L(t'',t'),\,\,\,\, \hat
L(t+T,t'+T)=\hat L(t,t').
\end{equation}

Using Eq.~(\ref{12}) together with Eq.~(\ref{50}), one can write
the average current on the form
\begin{equation}\label{13}
I=\frac{e}{T}\int_0^T
dt\Gamma_R(t)\int^t_{-\infty}dt'\Gamma_L(t')|\hat
L_{21}(t,t')|^2\,,
\end{equation}
where $\hat L_{21}(t,t')$ is a matrix element of the operator
$\hat L(t,t'); \hat L_{21}(t,t')= (\textbf{e}_2,\hat
L(t,t')\textbf{e}_1)$.

Since the probability amplitude for tunnelling is exponentially
sensitive to the position of the dot, the maximum of the tunnel
exchange interaction between an electron on the dot and an
electron in one lead occurs when the tunnelling coupling to the
other lead is negligible. This is why we will assume the following
property of tunnelling amplitude $\Lambda_{L,R}(t)$ to be
fulfilled:
\begin{equation}\label{02}
T_L(t)T_R(t)=0,\,\, T_L(t),T_R(t)\neq 0
\end{equation}
This assumption allows us to divide the time interval $(0,T)$ into
the intervals $(0,\tau)+(\tau,T/2)+(T/2,T/2+\tau)+(T/2+\tau,T)$.
We suppose that $T_L(t)\neq 0$ (but $H=0$) only in the time
interval $(0,\tau)$ (and, analogously, $T_R(t)\equiv
T_L(t+T/2)\neq 0$ in the time interval $(T/2,T/2+\tau)$). Using
this approximation together with the properties (\ref{120}) of the
operator $\hat L(t,t')$, we arrive at the following expression for
the average current (Appendix 2):
\begin{equation}\label{14}
I=\frac{e}{T}(1-e^{-\Gamma})^2 \sum_{n=0}^\infty |(\textbf{e} _2,
\hat L(T/2,\tau)\hat L^n \textbf{e}_1)|^2.
\end{equation}
Here $\hat L\equiv\hat L(T+\tau,\tau)$ and
\begin{equation}\label{140}
\Gamma=2 \pi \nu \int^\tau_0 dt T_L^2(t)
\end{equation}
is the tunnelling rate. Consequently, in order the calculate the
average current it is necessary to investigate the properties of
the evolution operator $\hat L$. It follows from its definition
that
\begin{eqnarray}\label{15}
\hat L&=& \hat L(T+\tau, \tau)= \hat L(T+\tau, T)\hat
L(T,T/2+\tau) \hat L(T/2+\tau,T/2)\hat L(T/2,\tau) \\&=&
e^{-(1+\sigma_3)\Gamma/4+i \sigma_3\Phi_0}\hat
L(T,T/2+\tau)e^{-(1-\sigma_3)\Gamma/4-i \sigma_3\Phi_0} \hat
L(T/2,\tau)\,,\nonumber
\end{eqnarray}
where $\Phi_0=\int_0^\tau dt J(t)\,$. From the symmetry properties
of the operator $\hat R(T/2+\tau<t<T)$, $$ \hat R^\dag=\hat R,
\sigma_2\hat{R}^\ast=-\hat R, \sigma_3\hat R(-t)=-\hat
R(t)\sigma_3$$ it follows that the operator $\hat U\equiv \hat
L(T,T/2+\tau)$ has the form
\begin{equation}\label{16}
\hat U=\left( \begin{array}{cc} \sqrt{1-\gamma^2} & i\gamma
e^{i\varphi} \\ i\gamma e^{-i\varphi} & \sqrt{1-\gamma^2}
\end{array}\right)
\end{equation}
In addition to this, $\hat L(T/2,\tau)=\sigma_1\hat U \sigma_1$.
As a result, the operator $\hat L$ can be expressed as
\begin{equation}\label{17}
\hat L = e^{-\Gamma/2}\left( e^{-\sigma_3\Gamma/4+i\Phi_0\sigma_3}
\hat U \sigma_1\right)^2\,.
\end{equation}

Proceeding with the analysis we (i) calculate the eigenvalues
$\lambda_{i}$ and eigenvectors $\textbf{b}_i$ of the operator
$\hat L$ of Eq.~(\ref{17}); $ \hat L\textbf{b}_i= \lambda_{i}
\textbf{b}_i$, (ii) substitute the expansion $\textbf{e}_i= a_{ji}
\textbf{b}_j$ (where $(a)^{-1} = (\textbf{e}_{i},\textbf{b}_{j}$))
into Eq.~(\ref{14}) and calculate the average current. The result
is
\begin{equation}\label{18}
I=\frac{e\kappa}{T}\sinh\Gamma/2\frac{\cosh\Gamma/2+\cos 2
\vartheta}{\sinh^2 \Gamma/2+\kappa(1+\cos
2\vartheta\cosh\Gamma/2)}\,,
\end{equation}
where $\vartheta=\varphi+\Phi_0, \kappa=2\gamma^2/(1+\gamma^2)$.
Equation~(\ref{18}) for the average current is our main result.

\section{Calculation of the Current in Limit of Strong and Weak Exchange
Coupling between the Dot and the Leads}

Although the result (\ref{18}) for the tunnel current is both
transparent and compact, it is in general a rather complicated
problem to find the magnetic field dependence of the coefficient
$\kappa$, which depends on the probability amplitude $\gamma$ for
flipping the spin of shuttled electrons. Three different time
scales are involved in the spin dynamics of a shuttled electron.
They correspond to three characteristic frequencies: (i) the
frequency of spin rotation, determined by the tunnel exchange
interaction with the magnetic leads; (ii) the frequency of spin
rotation in the external magnetic field, and (iii) the frequency
of shuttle vibrations. Different regimes occur depending on the
relation between these time scales. Here we will consider two
limiting cases, where a simple solution of the problem can be
found. Those are the limits of weak $J_{L(R)}\ll \mu H$ and strong
$J_{L(R)}\gg \mu H$ exchange interactions with the leads.

\subsection{Weak exchange interaction}
In the limit $ J_{L(R)}\ll \mu H $ one may neglect the influence
of the magnetic leads on the on-dot electron spin dynamics. In
this case the matrix $\hat U$ given by Eq.~(\ref{16}) can easily
be calculated and Eq.~(\ref{18}) reduces to
\begin{equation}\label{19}
I=\frac{2e}{T}\frac{\sin^2\vartheta/2\,
\tanh\Gamma/4}{\sin^2\vartheta/2+\tanh^2\Gamma/4}\,,
\end{equation}
where $\vartheta=g\mu\int_\tau^{T/2}dt H$ is the rotation angle of
the spin in the external field.

Two different scales for the external magnetic field determine the
magneto-transmittance in this limit. One scale is associated with
the width of the resonant magnetic field dependence (see the
denominator in Eq.~(\ref{19})). This scale is (restoring
dimension)
\begin{equation}\label{119}
\delta H = \Gamma\frac{ \hbar\omega}{g \mu}\,,
\end{equation}
where $\omega$ is the shuttle vibration frequency. The second
scale,
\begin{equation}\label{019}
\Delta H =\frac{\hbar\omega}{g\mu}\,,
\end{equation}
comes from the periodic function $ \sin^2\vartheta/2$ that enters
Eq.~(\ref{19}). The magnetic-field dependence of the current is
presented in Fig.~2a. Dips in the transmittance of width $\delta
H$ appear periodically as the magnetic field is varied, the period
being $\Delta$. This amount to a giant magneto-transmittance
effect. It is interesting to notice that by measuring the period
of the variations in $I(H)$ one can in principle determine the
shuttle vibration frequency. This amounts to a dc method for
spectroscopy of the nanomechanical vibrations.
Equation~(\ref{019}) gives a simple relation between the vibration
frequency and the period of the current variations. The physical
meaning of this relation is very simple: every time when
$\omega/\Omega=n+1/2$ ($\Omega$ is the spin precession frequency
in a magnetic field) the shuttled electron is able to fully flip
its spin to remove the "spin-blockade" of tunnelling between spin
polarized leads having their magnetization in opposite directions.

\begin{figure}
\vspace{-0.5cm} \centerline{\hspace{3cm}
\includegraphics[width=20cm]{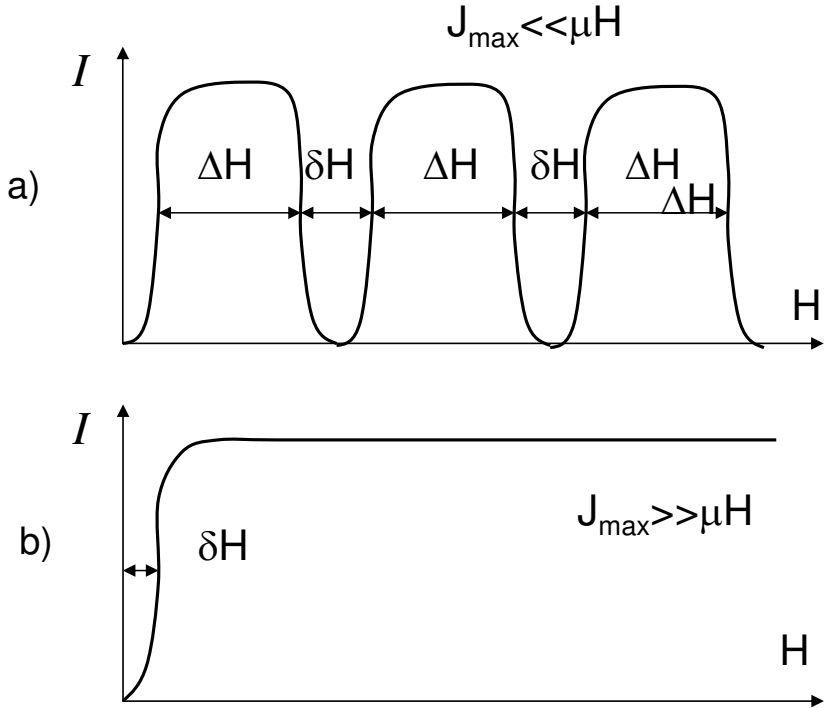} } \vspace*{-3cm}
\caption{Magnetic-field dependence of the transmittance of the
device shown in Fig.~1 for the limiting cases of a) weak and b)
strong exchange coupling between dot and leads. The period $\Delta
H$ and the width $\delta H$ of the "dips" are given by Eqs.~(21)
and (20) for case a) and $\delta H$ is given by Eq.~(24) for the
case b).}
\end{figure}

\subsection{Strong Dot-Leads  Exchange Interaction}

A strong magnetic coupling to the leads, $J_{\text{max}}\gg \mu
H$, preserves the electron spin polarization, preventing
spin-flips of shuttled electrons due to an external magnetic
field. However, if the magnetization of the two leads are in
opposite directions, the exchange coupling to the leads have
different sign. Therefore, the exchange couplings to the two leads
tend to cancel out when the dot is in the middle of the junction.
Hence the strong exchange interaction affecting a dot electron
depends on time and periodically changes sign, being arbitrary
small close to the time of sign reversal. In Fig.~3 the on-dot
electronic energy levels for spins parallel and antiparallel to
the lead magnetization are presented as a function of time. The
effect of an external magnetic field is in the limit $J_{L(R)}\gg
\mu H$ negligible almost everywhere, except in the vicinity of the
level crossing. At this "time point" ,which we denote $t_{LZ}$,
the external magnetic field removes the degeneracy and a gap is
formed in the spectrum (dashed curve). The probability of
electronic spin-flip in this case is determined by the probability
of a Landau-Zener reflection from the gap formed by the magnetic
field (in this case a Landau-Zener transition across the gap is a
mechanism for backscattering of the electron, since this is the
channel where the electronic spin is preserved). The matrix $\hat
U$ can readily be expressed in terms of Landau-Zener scattering
amplitudes. The amplitude and phase of electronic spin-flip is
given by $\varphi=\varphi_0+\Phi_1, \varphi_0$ is the Landau-Zener
phase shift,
\begin{equation} \label{543}
\Phi_1=\int_\tau^{T/2-\tau}dt J(t)
\end{equation}
and $\gamma^2$ is the probability of the Landau-Zener "backward"
scattering,
\begin{equation} \label{544}
\gamma^2=1-\exp\left[-\frac{\pi(\mu H)^2}{J'(t_{LZ})}\right]\,.
\end{equation}
Schematical view of $I(H)$ dependence is presented on a Fig.2b.
The width $\delta H$ of the minimum in $I(H)$ dependence can be
found directly from Eqs.(\ref{18}), (\ref{544})
\begin{equation}\label{555}
    \delta H = \frac{\pi g \mu}{\sqrt{J_0 \hbar \omega}}\,,
\end{equation}
where $J_0 = \text{min} (J_{L(R)}(t))$.

\begin{figure}
\vspace{-0.5cm} \centerline{\hspace{6cm}
\includegraphics[width=20cm]{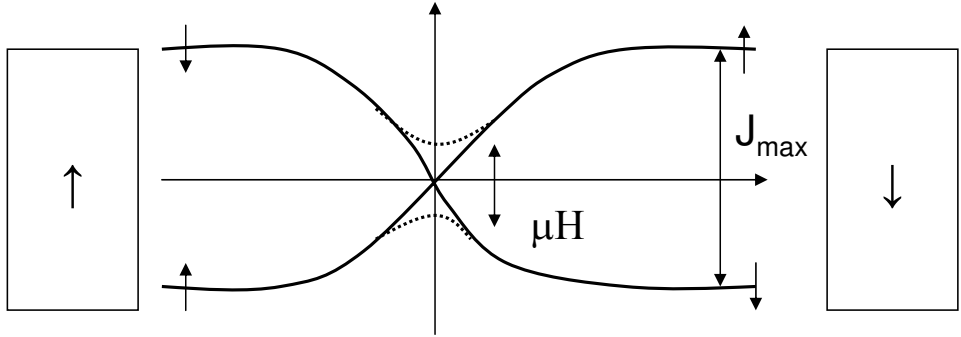} } \vspace*{-7cm}
\caption{On-dot energy levels for spin-up and spin-down electron
states as a function of the position of the dot. Level crossing in
the middle of the device is removed by an external magnetic
field.}
\end{figure}

\section{Conclusion}
The analysis presented above demonstrates the possibility of a
giant magneto-transmittance effect caused by shuttling of
spin-polarized electrons between magnetic source- and drain
electrodes. The sensitivity of the shuttle current to an external
magnetic field is determined, according to Eq.~(\ref{119}), by the
transparency of the tunnel barriers. By diminishing the tunnelling
transmittance one can increase the sensitivity of the device to an
external magnetic field. The necessity to have a measurable
current determines the limit of this sensitivity. In the low
transparency limit, $\Gamma\ll 1 $, the current through the device
can be estimated as $I\simeq e\Gamma\omega$. If one denotes the
critical field that determines the sensitivity of the device by
$H_{cr}$, one finds from Eq.~(\ref{119}) that $H_{cr}\simeq \delta
H$. The critical field can now be expressed in terms of the
current transmitted through the device as
\begin{equation}\label{224}
H_{cr}(\text{Oe})\simeq \frac{\hbar I}{e\mu g}\simeq \frac{g_0}{g}
(3\times 10^2) I(\text{n\AA})\,,
\end{equation}
where $g_0\,(=2)$ is the gyromagnetic ratio for the free
electrons. For $I\simeq 10^{-1} \div 10^{-2}$ nA and $g_0/g\simeq
1/3$ this gives a range $H_{cr}\simeq 1\div 10 \,\text{Oe}$. A
further increase in sensitivity would follow if one could use a
shuttle with several ($N$) electronic levels involved in the
tunnelling process. The critical magnetic field would then be
inversely proportional to the number of levels,
$H_{cr}(N)=H_{cr}(N=1)/N$.

\subsection{Acknowledgment}
Financial support from the Royal Swedish Academy of sciences
(SIK), the Swedish Science Research Council (LIG, RIS) and the
Swedish Foundation for Strategic Research (LIG, RIS, MJ) is
gratefully acknowledged. Support from the U.S.Department of energy
(Basic Energy Science, Contract No W-31-109-ENG-38) is gratefully
acknowledged.

\subsection{Appendix 1}
The Shr\"{o}dinger equation results in equations for the
coefficients $c^{\alpha\beta}_{R(L)},
c^\alpha_{\uparrow(\downarrow)}$:
\begin{eqnarray}\label{23}
i\frac{\partial c^\alpha_\uparrow}{\partial t}&=&
-J(t)c^\alpha_\uparrow- (g\mu H/2) c^\alpha_\downarrow + T_L(t)
\sum_\beta c^{\alpha\beta}_L (t)\,,\\ i\frac{\partial
c^\alpha_\downarrow}{\partial t}&=& J(t)c^\alpha_\uparrow- (g\mu
H/2) c^\alpha_\downarrow + T_R(t) \sum_\beta
c^{\alpha\beta}_R(t)\,,\nonumber \\ i\frac{\partial
c^{\alpha\beta}_L}{\partial t}&=&\varepsilon_\beta
c^{\alpha\beta}_L+ T_L(t) c^\alpha_\uparrow (t)\,, \nonumber \\
i\frac{\partial c^{\alpha\beta}_R}{\partial t}&=&\varepsilon_\beta
c^{\alpha\beta}_R+ T_R(t) c^\alpha_\downarrow(t)\,. \nonumber
\end{eqnarray}

As it follows from the last two equations (together with the
initial conditions)
\begin{eqnarray}\label{24}
c^{\alpha\beta}_L(t)& = &e^{-i \varepsilon_\beta t}
\delta_{\alpha\beta} - i \int_{-\infty}^t dt'
e^{i\varepsilon_\beta(t'-t)}T_L(t')c^\alpha_\uparrow(t')\,,
\\c^{\alpha\beta}_R(t)& = & - i \int_{-\infty}^t dt'
e^{i\varepsilon_\beta(t'-t)}T_R(t')c^\alpha_\downarrow(t')\,.
\nonumber
\end{eqnarray}
Therefore, for the $\sum_\beta c^{\alpha\beta}_R (t)$ one gets $$
\sum_\beta c^{\alpha\beta}_R (t)=-i\int_{-\infty}^t dt'T_R(t')
c^\alpha_\downarrow(t') \sum_\beta e^{i\varepsilon_\beta
(t'-t)}\,.$$ In wide-band approximation we suppose
$\nu(\varepsilon)=\text{const}$, therefore $ \sum_\beta
e^{i\varepsilon_\beta (t'-t)}=2\pi \nu \delta (t'-t)$ and
\begin{equation}\label{25}
\sum_\beta c^{\alpha\beta}_R (t)=-i \pi\nu
T_R(t)c^\alpha_\downarrow \,.
\end{equation}
Analogously,
\begin{equation}\label{26}
\sum_\beta c^{\alpha\beta}_L (t)=e^{-i\varepsilon_\alpha t}-i
\pi\nu T_L(t)c^\alpha_\uparrow \,.
\end{equation}
Substitute the expressions, Eqs.(\ref{25}), (\ref{26}), to the
first two equations (\ref{23}), one get the equation Eq.~(\ref{9})
for the bi-vector $\textbf{c}^{\alpha}$.

\subsection{Appendix 2}
Under our approximation we can  change the integration limits in
Eq.~(\ref{13}):
\begin{eqnarray}\label{a1}
I&=&(2\pi\nu)^2\frac{e}{T}\int_0^T dt T_R^2(t)
\int^t_{-\infty}dt'T_L^2(t')|\hat L_{21}(t,t')|^2 \nonumber \\&=&
(2\pi\nu)^2\frac{e}{T}\int_{T/2}^{T/2 +\tau} dt T_R^2(t)
\int^{\tau}_{-\infty} dt'T_L^2(t')|\hat L_{21}(t,t')|^2\,.
\end{eqnarray}
Beside this, in the time moments $T/2<t<T/2+\tau\, \hat L(t,T/2)$
is a diagonal matrix. Therefore $\hat L_{21}(t,t') = \hat
L_{22}(t,T/2) \hat L_{21}(T/2, t')$. As a consequence, the
integral in the expression for the average current,
Eq.~(\ref{a1}), is factorized:
\begin{equation}\label{a4}
I=(2\pi\nu)^2\frac{e}{T}\int_{T/2}^{T/2+\tau}  dt T_R^2(t) |\hat
L_{22} (t,T/2)|^2 \int^{\tau}_{-\infty} dt'T_L^2(t')|\hat
L_{21}(T/2,t')|^2\,.
\end{equation}
The first integral in Eq.~(\ref{a4}) is easy to calculate. Having
in mind that ($T/2<t<T/2+\tau$) $$|\hat L_{22}(t,T/2)|^2=\exp
\left[ -2\pi\nu \int_{T/2}^t dt T_R^2(t)\right]\,,$$ one gets
\begin{equation}\label{a5}
2\pi\nu\int_{T/2}^{T/2+\tau}  dt T_R^2(t) |\hat L_{22} (t,T/2)|^2
= 1-e^{-\Gamma}\,,
\end{equation}
where quantity $\Gamma$ is defined in Eq.~(\ref{140}).

The calculation of the second integral in Eq.~(\ref{a4}) can be
done in the same manner. One has the set of equalities,
\begin{eqnarray}\label{a6}
&&\int_{-\infty}^\tau T_L^2(t)|\hat L_{21}(T/2,t)|^2=
\sum_{n=0}^\infty \int_{-nT}^{-nT+\tau} dt T_L^2(t)|\hat
L_{21}(T/2,t)|^2 \nonumber \\&&= \sum_{n=0}^\infty \int_0^\tau dt
T_L^2(t)|( \textbf{e}_2 ,\hat L (T/2, \tau) \hat L (\tau,
t-nT)\textbf{e}_1)|^2\,.
\end{eqnarray}
For the quantity $\hat L(\tau, t-nT)= \hat L(\tau+nT, t) $ one has
\begin{equation}\label{a7}
 \hat L(\tau +nT, t)=\hat L(\tau +nT, \tau +(n-1)T)\hat L(\tau
+(n-1)T, \tau +(n-2)T)...\hat L(\tau,t) = \hat L^n
(\tau+T,\tau)\hat L(\tau,t)\,.
\end{equation}

Therefore,
\begin{equation}\label{a8}
\int_{-\infty}^{\tau} dt T_L^2(t)|\hat L_{21}(T/2,t)|^2=
\sum_{n=0}^\infty \int_0^\tau dt T_L^2(t) |( \textbf{e}_2,\hat L
(T/2, \tau) \hat L^n \hat L(\tau,t) \textbf{e}_1)|^2\,.
\end{equation}
Calculating  the integral in the same manner, as in
Eq.~(\ref{a5}), one gets the Eq.~(\ref{14}) for the average
current.

\end{document}